\documentstyle[11pt,moriond,epsfig]{article}

\def\Journal#1#2#3#4{{#1} {\bf #2}, #3 (#4)}

\def\be{\begin{equation}}
\def\ee{\end{equation}}
\def\bea{\begin{eqnarray}}
\def\eea{\end{eqnarray}}

\begin{document}
\vspace*{4cm}
\title{COALESCENCE RATES OF DOUBLE NEUTRON STARS}

\author{ V. KALOGERA}

\address{Harvard-Smithsonian Center for Astrophysics, 60 Garden St.,
Cambridge, MA 02138, USA}

\maketitle

\abstracts{Merger events of close double neutron stars (DNS) lie at the
basis of a number of current issues in relativistic astrophysics, such as
the indirect and possible direct detection of gravitational waves, the
production of gamma-ray bursts at cosmological distances, and the origin
of r-process elements in the universe. In assessing the importance or
relevance of DNS coalescence to these issues, knowledge of the rate of
coalescence in our Galaxy is required.  In this paper, I review the
current estimates of the DNS merger rate (theoretical and empirical) and
discuss new ways to obtain limits on this rate using {\em all} information
available at present.}

\section{Introduction}

Double neutron star (DNS) systems observed as binary radio pulsars have
provided striking empirical confirmation of general relativity with the
measurement of orbital decay accompanied with gravitational-wave emission
in the prototypical system PSR B1913+16~\cite{TW82}. This orbital period
decrease is expected to lead eventually to the merger of the two neutron
stars. DNS coalescence events represent one of the most promising sources
for the {\em direct} detection of gravitational waves by the new
generation of laser-interferometer detectors currently under
construction~\cite{A92} (e.g., LIGO, VIRGO). The final coalescence of DNS
has also been discussed as a possible central engine of gamma-ray
bursts~\cite{P86}$^{,}$\cite{R97} and has been suggested as the possibly
dominant source of r-process elements in the universe~\cite{R99}.

Estimates of the rate of DNS coalescence in our Galaxy are crucial for
assessing the prospects for gravitational-wave detection and the possible
connection of DNS mergers to gamma-ray bursts and r-process elements.
Formation rates of {\em coalescing} DNS (systems with tight enough orbits
that merge within a Hubble time) have been calculated so far either
theoretically, based on evolutionary models of DNS formation, or
empirically, based on the observed DNS sample. In this paper I present a
critical review of the current coalescence-rate estimates addressing the
uncertainties involved and I discuss new ways of constraining these
estimates using all available observational information and current
theoretical understanding. It is useful to provide a scale for the various
estimates in the context of gravitational-wave detection: based on the
expected performance of the second-generation LIGO
observatories~\cite{T96}, for a detection rate of one merger event per
year, a Galactic DNS coalescence rate of $\sim 10^{-5}$\,yr$^{-1}$ is
required. For comparison, the estimated~\cite{W98} rate of gamma-ray
bursts per galaxy is $\sim 10^{-7}$\,yr$^{-1}$.

\section{Theoretical Estimates}

Theoretical calculations of the formation rate of coalescing DNS are
possible, given a sequence of evolutionary stages that leads from
primordial binaries to DNS formation. Over the years a relatively standard
picture has been formed describing the birth of DNS~\cite{H76}, although
more recently variations of it have also been discussed~\cite{B95}. In all
versions of the DNS formation path the main picture remains the same. The
initial binary progenitor consists of two binary members massive enough to
eventually collapse into neutron stars. Its evolution involves multiple
phases of stable or unstable mass transfer, common-envelope evolution, and
accretion onto neutron stars, as well as two supernova explosions.

Such theoretical modeling of DNS formation has been undertaken by various
authors by means of population syntheses. This provides us with {\em ab
initio} predictions of the coalescence rate. The evolution of an ensemble
of primordial binaries with assumed initial properties is followed through
specific evolutionary stages until a coalescing DNS is formed. The changes
in the properties of the binaries at the end of each stage are calculated
based on our current understanding of the various processes involved: wind
mass loss from massive hydrogen- and helium-rich stars, mass and
angular-momentum losses during mass transfer phases, dynamically unstable
mass transfer and common-envelope evolution, effects of highly
super-Eddington accretion onto neutron stars, and supernova explosions
with kicks imparted to newborn neutron stars. Given that several of these
phases are not very well understood, the results of population synthesis
are expected to depend on the assumptions made in the treatment of the
various processes. Therefore, exhaustive parameter studies are required by
the nature of the problem.

Recent studies of DNS formation and calculations of coalescence
rates~\cite{L97}$^{,}$\cite{F98}$^{,}$\cite{PZ98}$^{,}$\cite{BB98}$^{,}$
\cite{BB99} have explored the input parameter space and the robustness of
the results at different levels of (in)completeness. Almost all have
studied the sensitivity of the coalescence rate to the average magnitude
of the kicks imparted to newborn neutron stars. The range of predicted
Galactic rates from {\em all} these studies obtained by varying the kick
magnitude is $5\times 10^{-7}~-~5\times 10^{-4}$\,yr$^{-1}$. This large
range indicates the importance of supernovae (two in this case) in
binaries. Variations in the assumed mass-ratio distribution for the
primordial binaries can {\em further} change the predicted rate by about a
factor of $10$, while assumptions of the common-envelope phase add another
factor of about $10-100$. Variation in other parameters typically affects
the results by factors of two or less.

It is evident that recent theoretical predictions for the DNS coalescence
rate cover a disappointingly wide range of values (typically 3-4 orders of
magnitude), which actually includes the ``nominal'' value of $\sim
10^{-5}$\,yr$^{-1}$. Note that DNS properties other than the coalescence
rate, such as orbital sizes, eccentricities, center-of-mass velocities,
are much less sensitive to the various input parameters and assumptions;
the latter affect more severely the absolute normalization (birth rate) of
the population. Given these results it seems fair to say that population
synthesis calculations have very limited predictive power and provide
fairly loose constraints on the DNS coalescence rate. Overall, we cannot
use them to make a robust statement about the prospects for detection of
merger events by the upcoming gravitational-wave observatories.

\section{Empirical Estimates}

Another way to estimate the coalescence rate is to use the properties of
the observed coalescing DNS (only two systems: PSR B1913+16 and PSR
B1534+12) combined with models of selection effects in radio pulsar
surveys. For each observed object, a scale factor is calculated based on
the fraction of the Galactic volume within which pulsars with properties
identical to those of the observed pulsar could be detected, in principle, 
by any of the radio pulsar surveys, given their detection thresholds. This
scale factor is a measure of how many more pulsars like those detected in
the coalescing DNS systems exist in our galaxy. The coalescence rate can
then be calculated based on the scale factors and estimates of detection
lifetimes summed up for all the observed systems.  This basic method was
first used by Phinney~\cite{P91} and Narayan et al.~\cite{N91} who
estimated the Galactic rate to be $\sim 1-3\times 10^{-5}$\,yr$^{-1}$. 

Since then, estimates of the coalescence rate have decreased significantly
primarily because of (i) the increase of the Galactic volume covered by
radio pulsar surveys with no additional coalescing DNS being
discovered~\cite{CL95}, (ii) the increase of the distance estimate for PSR
B1534+12 based on measurements of post-Newtonian parameters~\cite{S98}
(see also contribution by I.\ Stairs, this volume), (iii) changes in the
lifetime estimates for the observed systems~\cite{HL96}$^{,}$\cite{A99}.
The most recently published study~\cite{A99} gives a lower limit of
$2\times 10^{-7}$\,yr$^{-1}$ and a ``best'' estimate of $\sim 6-10\times
10^{-7}$\,yr$^{-1}$.

Some of the assumptions made in obtaining the above estimates, are not
clearly justifiable or testable. In particular, one assumes that the
sample of the two observed coalescing DNS is representative of the total
Galactic population~\cite{K99}, that the detection volume for each object
and its lifetime are independent and separable (see contribution by T.A.\
Prince, this volume, for arguments against this in the presence of pulsar
luminosity evolution and subsequent further reduction of the estimated
rate), and that the DNS pulsar luminosity function is similar to that of
young, non-recycled pulsars. Additional uncertainties arise from estimates
of pulsar ages and distances, the pulsar beaming fraction, the spatial
distribution of DNS in the Galaxy, and the number of undetectable pulsars
with luminosities below the detection limits of surveys.

Despite all these uncertainties the empirical estimates of the DNS
coalescence rate appear to span a range of $\sim 1-2$ orders of magnitude,
which is relatively narrow compared to the range covered by the
theoretical estimates.
 
\subsection{Small-number Sample}

One important limitation of empirical estimates of the coalescence rates
is that they are derived based on {\em only two} observed DNS systems,
under the assumption that the observed sample is representative of the
true population. Therefore, assessing the effect of small-number
statistics on the results of the above studies is necessary. Assuming that
DNS pulsars follow the radio luminosity function of young pulsars and that
therefore their population is dominated in number by low-luminosity
pulsars, it can be shown that the current empirical estimates
underestimate the true coalescence rate.  If a small-number sample is
drawn from a parent population dominated by low-luminosity (hence hard to
detect) objects, it is statistically more probable that the sample will
actually be dominated by objects from the high-luminosity end of the
population. Consequently, the empirical estimates based on such a sample
will tend to overestimate the detection volume for each observed system,
and therefore underestimate the scale factors and the resulting
coalescence rate (this effect is partly related to the Malmquist bias).

This effect can be clearly demonstrated with a Monte Carlo experiment (R.
Narayan, private communication) with the use of simple models for the
pulsar luminosity function and the survey selection effects. As a first
step, the average observed number of pulsars is calculated given a known
true total number of pulsars in the Galaxy (thick-solid line in Figure 1).
As a second step, a large number of sets of a given number of `observed'
(simulated) pulsars with assigned luminosities according to the assumed
luminosity function are realized using Monte Carlo methods. Based on each
of these sets, one can estimate the total number of pulsars in the Galaxy
using empirical scale factors, as is done for the real observed sample.
The many (simulated) `observed' samples can then be used to obtain the
distribution of the estimated total Galactic numbers of pulsars. The
median and 25\% and 75\% percentiles of this distribution are plotted as a
function of the assumed number of systems in the (fake) `observed' samples
in Figure 1 (thin-solid and dashed lines, respectively).

\begin{figure}
\centerline{\epsfig{figure=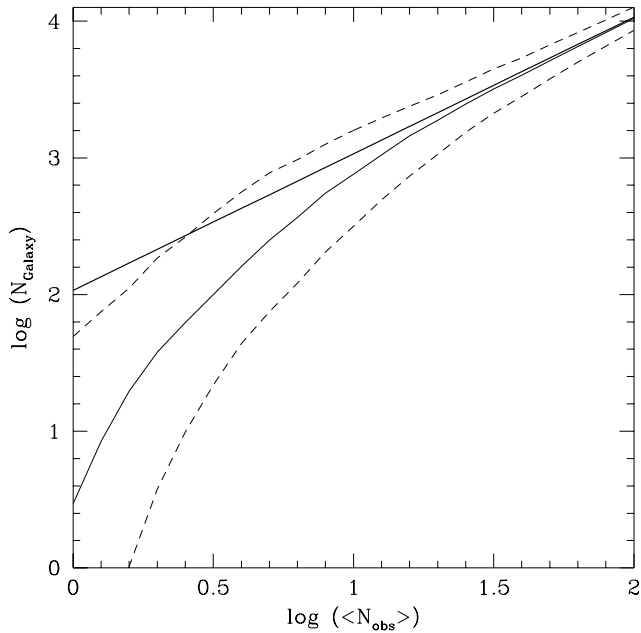,height=3.5in}}
\caption{Bias of the empirical estimates of the DNS coalescence rate because
of the small-number observed sample. See text for details.}
\end{figure}

It is evident from Figure 1 that, in the case of small-number observed
samples (less than $\sim 10$ objects), it is highly probably that the
estimated total number, and hence the estimated coalescence rate, is
underestimated by a significant factor. For a two-object sample, for
example, the true rate maybe higher by more than a factor of ten.

\section{Limits on Coalescence Rates} 

One way to circumvent the uncertainties involved in the estimates of the
DNS coalescence rate is to focus on obtaining upper or lower limits to
this rate. Depending on how their value compares to the value of $\sim
10^{-5}$\,yr$^{-1}$ needed for one LIGO II event per year, such limits can
provide us with valuable information about the prospects of
gravitational-wave detection.

Bailes~\cite{B96} used the absence of any young pulsars detected in DNS
systems and obtained a rough upper limit to the rate of $\sim
10^{-5}$\,yr$^{-1}$, while recently Arzoumanian {\it et al}~\cite{A99}
reexamined this in more detail and claimed a more robust upper limit of
$\sim 10^{-4}$\,yr$^{-1}$.

An upper bound to the rate can also be obtained by combining our
theoretical understanding of orbital dynamics (for supernovae with
neutron-star kicks occurring in binaries)  with empirical estimates of the
birth rates of {\em other} types of pulsars related to DNS
formation~\cite{KLN99}. Binary progenitors of DNS systems experience two
supernova explosions when the neutron stars are formed. The second
supernova explosion (forming the neutron star that is {\em not} observed
as a pulsar) provides a unique tool for the study of DNS formation, since
the post-supernova evolution of the system is simple, driven only by
gravitational-wave radiation.

There are three possible outcomes after the second supernova: (i) a
coalescing DNS is formed (CB), (ii) a wide DNS (with a coalescence time
longer than the Hubble time) is formed (WB), or (iii) the binary is
disrupted (D) and a single pulsar similar to the ones seen in DNS
(DNS-like) is ejected. Based on supernova orbital dynamics we can
calculate the probability branching ratios for these three outcomes,
$P_{\rm CB}$, $P_{\rm WB}$, and $P_{\rm D}$. For a given kick magnitude,
we can calculate the maximum ratio $(P_{\rm CB}/P_{\rm D})^{\rm max}$ for
the complete range of pre-supernova parameters defined by the necessary
constraint $P_{\rm CB}\neq 0$ (Figure 2). Given that the two types of
systems have a common parent progenitor population, the ratio of
probabilities is equal to the ratio of the birth rates $(BR_{\rm
CB}/BR_{\rm D})$.

\begin{figure}
\centerline{\epsfig{figure=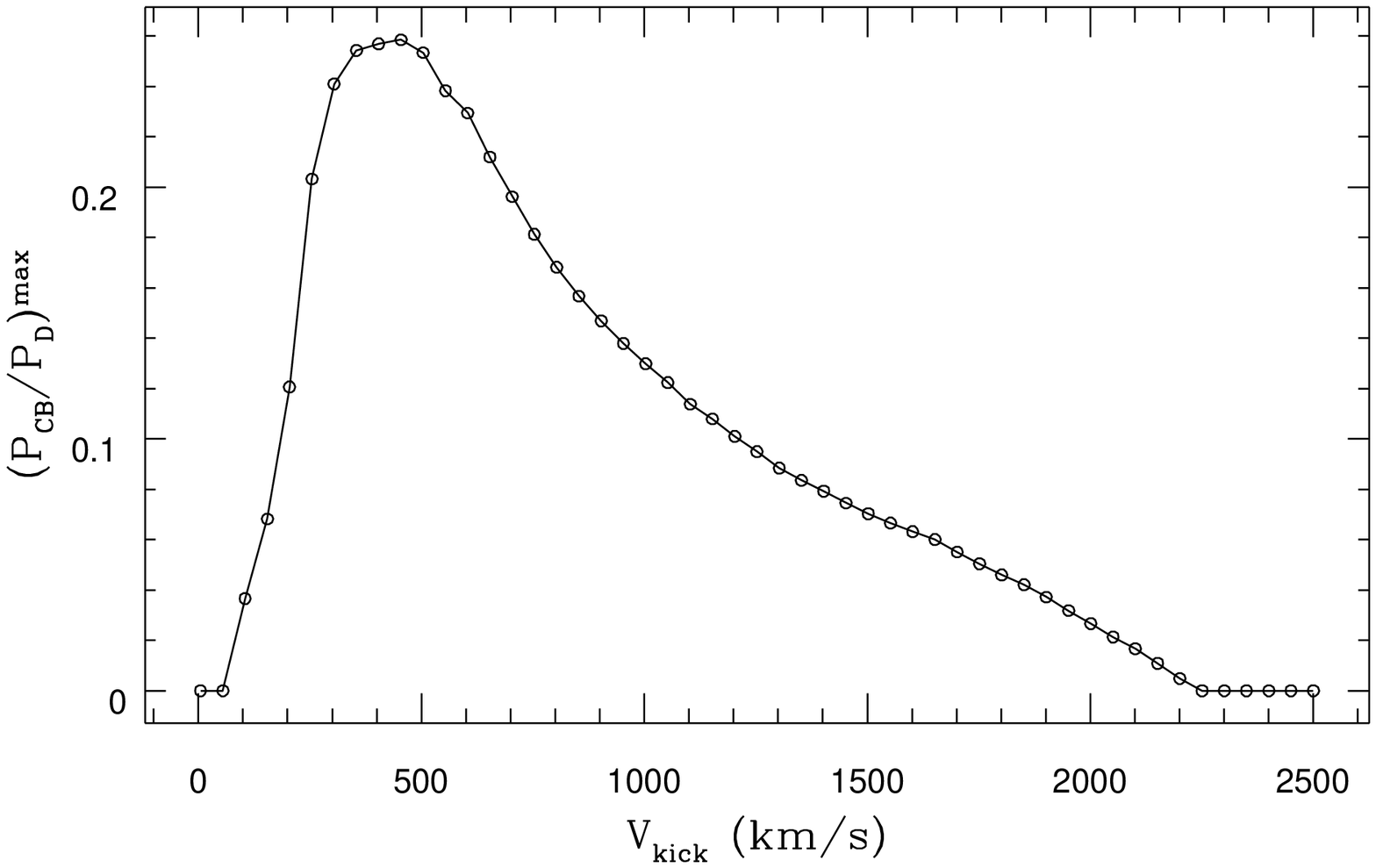,height=3.45in}}
\caption{Maximum probability ratio for the formation of coalescing DNS and the
disruption of binaries as a function of the kick magnitude at the second
supernova.}
\end{figure}

We can then use (i) the absolute maximum of the probability ratio
($\approx 0.26$ from Figure 2) and (ii) an empirical estimate of the birth
rate of single DNS-like pulsars based on the current observed sample to
obtain an upper limit to the DNS coalescence rate. The selection of this
small-number sample involves some subtleties~\cite{KLN99}, but a
preliminary analysis shows $BR_{\rm CB} > 0.5-3\times
10^{-5}$\,yr$^{-1}$~\cite{KLN99}. Note that this number could be increased
because of the small-number sample bias affecting this time the empirical
estimate of $BR_{\rm D}$ (see \S\,3.1).

This is an example of how we can use observed systems other than DNS to
improve our understanding of their coalescence rate. A similar calculation can
be done using the wide DNS systems instead of the single DNS-like
pulsars~\cite{KLN99}. 

\section{Conclusions}

A comparison of the various results on the DNS coalescence rate indicates
that theoretical estimates based on modeling of DNS formation have a quite
limited predictive power (range of at least 3-4 orders of magnitude),
whereas empirical estimates based on the observed DNS sample appear to be
more robust (within a factor of $\sim 100$).  Nevertheless, current
estimates, which fall right around the range of interest for evaluating
the prospects of gravitational-wave detection, still suffer from
uncertainties and systematic effects (e.g., small-number statistics,
distances, luminosity function, beaming). Therefore, the need for
additional constraints and alternative methods for estimates is clear. As
an example, a fairly robust upper bound to the rate can be obtained making
use all of the available information for other systems evolutionarily
linked to DNS.

Apart from DNS systems, binaries with two black-holes or a black-hole and
a neutron star are also of interest in the context of gravitational-wave
detection and possibly gamma-ray bursts. Since at present there are {\em
no} observed systems of this type, we have to rely on theoretically
predicted formation rates for them, keeping in mind the normalization
uncertainties associated with them. We note, however, that the absence of
such systems from the observed samples combined with some basic
understanding of their formation relative to DNS could provide valuable
information for the frequency of their formation in our Galaxy.

\section*{Acknowledgments}

I would like to thank T.A.\ Prince for discussions, R.\ Narayan for
discussions and for providing me with the results of his Monte Carlo
simulations, and D.R.\ Lorimer for providing me with an empirical estimate
of the birth rate of single DNS-like pulsars. I would also like to thank
D.\ Psaltis and F.\ Rasio for comments on an initial manuscript. I am
grateful to the organizers for inviting me to the meeting and acknowledge
full support by the Smithsonian Institute in the form of a CfA
Post-doctoral Fellowship.


\section*{References}
\begin{thebibliography}{99}

\bibitem{TW82} J.H. Taylor and J.M. Weisberg, \Journal{ApJ}{253}{908}{1982}.

\bibitem{A92} A. Abramovici {\it et al}, \Journal{Science}{256}{325}{1992}.

\bibitem{P86} B. Paczynski, \Journal{ApJ}{308}{L43}{1986}.

\bibitem{R97} M. Ruffert {\it et al}, \Journal{A\&A}{319}{122}{1997}.

\bibitem{R99} S. Rosswog {\it et al}, \Journal{A\&A}{341}{499}{1999}.

\bibitem{T96} K.S. Thorne in {\em Compact Stars in Binaries}, IAU Symp.
No. 165, eds. J. van Paradijs, E. P. J. van den Heuvel and E. Kuulkers (Kluwer
Academic Publishers, Dordrecht, 1996). 

\bibitem{W98} R.A.M.J. Wijers {\it et al}, \Journal{ApJ}{}{L}{1998}

\bibitem{H76} E.P.J. van den Heuvel in {\em Structure and Evolution of Close
Binary Systems}, IAU Symp. No. 73, eds. P. Eggleton, S. Mitton and J. Whelan
(Kluwer Academic Publishers, Dordrecht, 1976).

\bibitem{B95} G.E. Brown, \Journal{ApJ}{440}{270}{1995}.

\bibitem{L97} V.M. Lipunov {\it et al}, \Journal{MNRAS}{288}{245}{1997}.

\bibitem{F98} C.L. Fryer {\it et al}, \Journal{ApJ}{496}{333}{1998}.

\bibitem{PZ98} S.F. Portegies-Zwart and L.R. Yungel'son,
\Journal{A\&A}{332}{173}{1998}.

\bibitem{BB98} G.E. Brown and H. Bethe, \Journal{ApJ}{506}{780}{1998}

\bibitem{BB99} K. Belczynski and T. Bulik,
\Journal{A\&A}{submitted}{astro-ph/9901193}{1999}.

\bibitem{P91} E.S. Phinney, \Journal{ApJ}{380}{17}{1991}

\bibitem{N91} R. Narayan {\it et al}, \Journal{ApJ}{379}{17}{1991}

\bibitem{CL95} S.J. Curran and D.R. Lorimer, \Journal{MNRAS}{276}{347}{1995}

\bibitem{S98} I.H. Stairs {\it et al}, \Journal{ApJ}{505}{352}{1998}

\bibitem{HL96} E.P.J. van den Heuvel and D.R. Lorimer,
\Journal{MNRAS}{283}{37}{1996}

\bibitem{A99} Z. Arzoumanian {\it et al},
\Journal{ApJ}{submitted}{astro-ph/9811323}{1999}

\bibitem{K99} V. Kalogera, in preparation

\bibitem{B96} M. Bailes {\em Compact Stars in Binaries}, IAU Symp.
No. 165, eds. J. van Paradijs, E. P. J. van den Heuvel and E. Kuulkers (Kluwer
Academic Publishers, Dordrecht, 1996).

\bibitem{KLN99} V. Kalogera {\it et al}, in preparation

\end{thebibliography}
\end{document}